\documentclass{aastex631}

\begin{document}

\title{An Image Quality Evaluation and Masking Algorithm Based On Pre-trained Deep Neural Networks}

\correspondingauthor{Peng Jia}
\email{robinmartin20@gmail.com}

\author{Peng Jia}
\affiliation{College of Electronic Information and Optical Engineering, Taiyuan University of Technology, Taiyuan, 030024, China}

\author{Yu Song}
\affiliation{College of Electronic Information and Optical Engineering, Taiyuan University of Technology, Taiyuan, 030024, China}

\author{Jiameng Lv}
\affiliation{College of Electronic Information and Optical Engineering, Taiyuan University of Technology, Taiyuan, 030024, China}

\author{Runyu Ning}
\affiliation{College of Electronic Information and Optical Engineering, Taiyuan University of Technology, Taiyuan, 030024, China}



\begin{abstract}
With the growing amount of astronomical data, there is an increasing need for automated data processing pipelines, which can extract scientific information from observation data without human interventions. A critical aspect of these pipelines is the image quality evaluation and masking algorithm, which evaluates image qualities based on various factors such as cloud coverage, sky brightness, scattering light from the optical system, point spread function size and shape, and read-out noise. Occasionally, the algorithm requires masking of areas severely affected by noise. However, the algorithm often necessitates significant human interventions, reducing data processing efficiency. In this study, we present a deep learning based image quality evaluation algorithm that uses an autoencoder to learn features of high quality astronomical images. The trained autoencoder enables automatic evaluation of image quality and masking of noise affected areas. We have evaluated the performance of our algorithm using two test cases: images with point spread functions of varying full width half magnitude, and images with complex backgrounds. In the first scenario, our algorithm could effectively identify variations of the point spread functions, which can provide valuable reference information for photometry. In the second scenario, our method could successfully mask regions affected by complex regions, which could significantly increase the photometry accuracy. Our algorithm can be employed to automatically evaluate image quality obtained by different sky surveying projects, further increasing the speed and robustness of data processing pipelines.
\end{abstract}



\section{Introduction} \label{sec:intro}
The large amount of astronomical data has created an urgent need for automated data processing pipelines that can extract scientific information from observation data. As observation data can be easily affected by various factors such as sky background, atmospheric turbulence, and detector noise, it is important to assess the quality of the data and allocate different processing strategies accordingly. For astronomical images captured by optical telescopes with a wide field of view, the first step in data processing pipelines is to evaluate the image quality. Scientists will check if the images are affected by factors that could impact a significant portion of the images, such as clouds, sky background noise, CCD blooming and some other effects. Meanwhile, scientists will also check the point spread functions (PSFs) in the images to ensure that the telescopes are working properly. Based on the results obtained from these steps, images with PSFs of reasonable shape and size and adequate signal to noise ratio (SNR) will be sent to target detection algorithms for further processing \citep{bertin1996sextractor, burke2019deblending,turpin2020vetting,jia2020detection,sun2023pnet}.\\

Sometimes, parts of images that are significantly affected by noise with large spatial scales need to be masked, and the rest of these images will then be sent to target detection algorithms. It is also important to note that images with irregular PSFs will typically be sent to the telescope maintenance group for possible malfeasance analysis \citep{wang2018automated,nie2021constraining,zhan2022database,liaudat2023rethinking}. Since there are numerous factors that can affect image quality, building an image quality evaluation method is difficult, and human intervention is necessary to oversee the image quality evaluation results. Nowadays, time domain astronomy observations are expected to produce several terabytes of images each night \citep{liu2021sitian, law2022low, ofek2023large}, and given the urgent need for timely follow-up observations of certain transients, it would be impractical to manually inspect and mask each of these images.\\

Assessing the quality of images is a long-standing issue within the astronomical image processing community, and a variety of algorithms have been proposed to tackle this problem. These algorithms can be broadly classified into three categories: full reference (FR) image quality evaluation, no-reference (NR) image quality evaluation, and reduced reference (RR) image quality evaluation. Among these categories, the FR approach has been extensively discussed in the literature, with the Peak Signal to Noise Ratio (PSNR) and the Structural Similarity (SSIM) being the most commonly used metrics \citep{SSIM}. These methods compare a reference image with the image being evaluated, and if the reference image has a high resolution and SNR, images with better quality will yield higher PSNR or SSIM scores. FR is a valuable method for evaluating the performance of image restoration algorithms \citep{schawinski2017generative,Ramos2018,jia2021data,ramos2021learning} and images that are similar \citep{Denker2005}, provided that appropriate reference images are available.\\

Because reference images are not possible to obtain in real applications, scientists have proposed various NR approaches to evaluate images without relying on high-resolution reference images \citep{NR}. The NR approach involves a set of manually designed filters that assess image quality based on their features. Solar images are typically evaluated using the NR approach due to their similar structures and being primarily affected by atmospheric turbulence-induced PSFs \citep{Deng2007, Denker2007, Popowicz2017}. However, design and testing of different NR algorithms manually for images obtained through time domain astronomy observations is not practical due to the large volume and highly variable properties of the data. As deep neural networks (DNN) now enable the modeling of human experience \citep{LeCun2015}, the emergence of DNNs offers the possibility of extending the no-reference (NR) approach, which is fundamentally developed based on human experience, to the reduced reference (RR) approach.\\

The RR approach involves the selection of specific images as references and utilizing DNNs to learn their features \citep{RR}. These learnt features can be used to evaluate the quality of other images. \citet{2020Assessment} have developed an automatic image quality evaluation network that combines clustering and Convolutional Neural Network (CNN) to classify images with different qualities. Additionally, \citet{Gram_matrix} have proposed a method for evaluating image quality by extracting image features using CNNs and calculating the Gram matrix between the extracted features of reference images and those of the images to be evaluated. However, selecting adequate reference images for sky surveying projects is a challenge in these RR approaches. \\

The concept of large models (LM) has garnered significant attention in recent years. Two prominent examples are the large language model (LLM) \citep{radford2019language, Thoppilan2022} and the large vision model (LVM) \citep{Kirillov2023}. These models serve as robust pretraining models that can be directly applied to various downstream applications once appropriate interfaces are designed. However, training LMs is a challenging task. While a substantial amount of data can be collected, a significant portion of it remains unlabeled. Therefore, unsupervised learning, referred to as pre-training, has become a crucial aspect of LM training \citep{brown2020language,2022arXiv220414198A}. Various pre-training methods have been developed for LLMs, including masked-language modeling, next sentence prediction, clustering, and knowledge distillation \citep{2019arXiv190109960H}. Similarly, analogous frameworks have been proposed for LVMs \citep{dosovitskiy2021image,jia2021scaling}. The pre-training process effectively incorporates prior information into LMs automatically. In the context of image quality evaluation, since human scientists also rely on prior information to assess image quality, it is conceivable to adapt a pre-training model for the purpose of evaluating image quality.\\

We propose our image quality evaluation algorithm based on this idea. To begin with, for images collected by a specific observation projects, we use a large dataset of high-resolution images with adequate SNRs as reference images and train an autoencoder in an unsupervised manner, which is similar to the pre-training stage for LMs. After that, we employ the pre-trained neural network to reconstruct observed images and calculate the difference between the reconstructed observed image and the original image. Based on the difference between these two images, we can obtain scores for the observed images. Images with larger score mean they are more different than images in the space defined by images with adequate SNRs. By splitting the images into small patches, we can effectively mask the parts of the images that are seriously affected by noise based on their scores. We test the performance of our algorithm using both simulated and real observation images, and results show that our algorithm is effective. This paper is structured as follows: In Section 2, we introduce the main structure of our algorithm. In Section 3, we test the performance of our algorithm with simulated observation images and real observational data in two different observation scenarios. We present our conclusions and outlook for future works in Section 4.\\

\section{Method} \label{sec:Method}
Our algorithm consists of two components: a DNN and a framework that utilizes the DNN to evaluate and mask parts of images. The DNN is trained on high-quality images to learn their features and can then reconstruct images within the feature space. By comparing the reconstructed images to the original images, we can evaluate the quality of the original images. The framework divides input images into smaller patches and evaluates the quality of each patch using the DNN. If the quality score of the patch falls below a certain criterion, the framework will mask that part of the image. We will discuss the specifics of our algorithm in the following section.\\

\subsection{The Structure of the DNN for Image Quality Evaluation} \label{subsec:nn}
\begin{figure} [ht]
   \begin{center}
   \begin{tabular}{c} 
   \includegraphics[height=5cm]{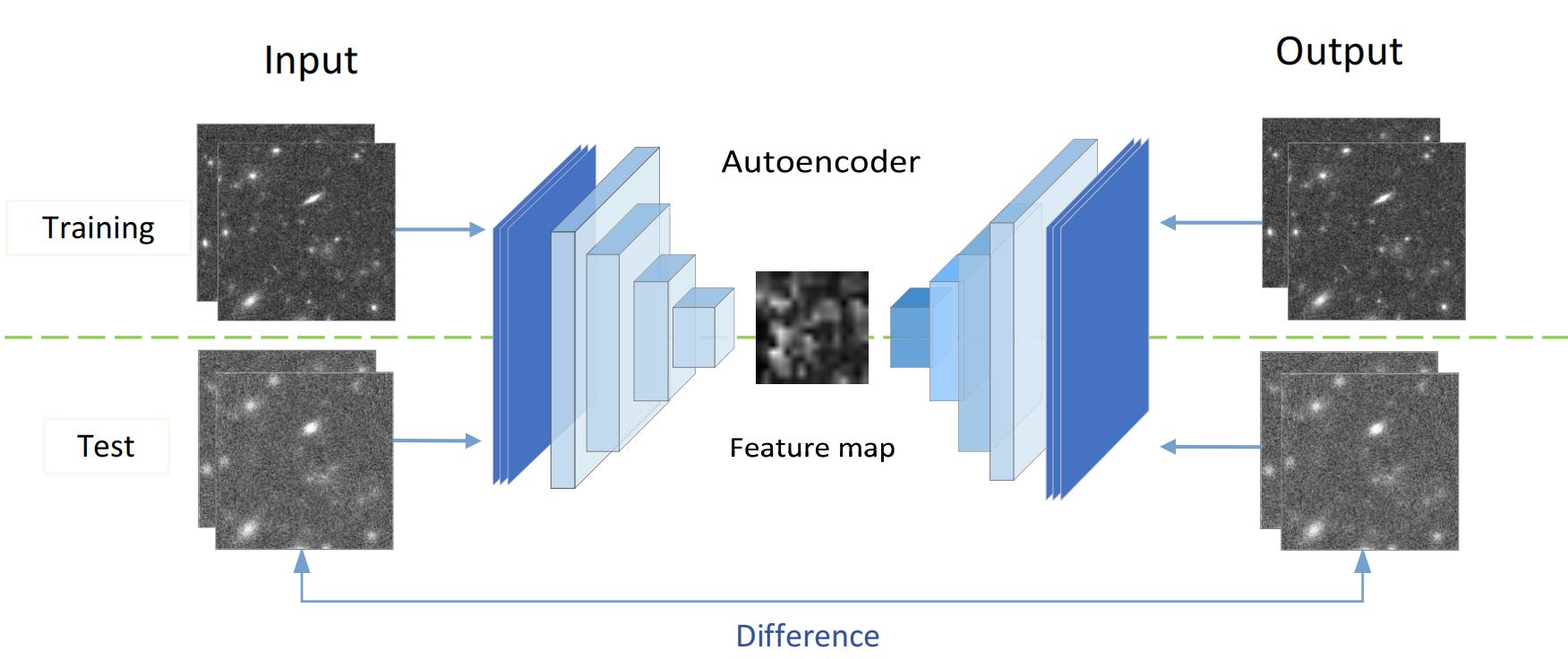}
   \end{tabular}
   \end{center}
   \caption{The figure illustrates the structure of the autoencoder, which first encodes input images into latent vectors and subsequently decodes the latent vectors to output images.
   \label{model structure}}
   \end{figure}

As previously discussed, the DNN is utilized to reconstruct images using features learned from high-quality images. By comparing the original images with their reconstructed counterparts, the quality of these images can be assessed \citep{LIU2014844,10.1007/978-3-030-03801-4_34}. To implement this approach, it is essential to choose a DNN that can effectively represent images with intricate details and can be trained in an unsupervised manner. In this paper, the autoencoder is selected as the prototype structure of the DNN, as depicted in Figure \ref{model structure}.\\

 An autoencoder is an unsupervised machine learning algorithm that learns to reconstruct input data in a compressed feature space \citep{dendi2018generating,yang2018blind,tripathi2021facial,hou2017deep}. Initially, the encoder compresses the input data into a lower-dimension representation, known as the latent space. Subsequently, the decoder reconstructs the images from vectors in the latent space. During the training phase, the autoencoder attempts to minimize the difference between the input images and the output images based on mean square error. After training, the encoder will learn a meaningful representation of the input data, while the decoder will learn how to reconstruct the original data from the representation. If an image is not similar to the images in the training set, there will be a discrepancy between the input and reconstructed images. Based on this idea, the autoencoder is trained with high-quality images and utilized to evaluate images with different qualities.\\
 
\begin{figure} [ht]
   \begin{center}
   \begin{tabular}{c} 
   \includegraphics[height=8cm]{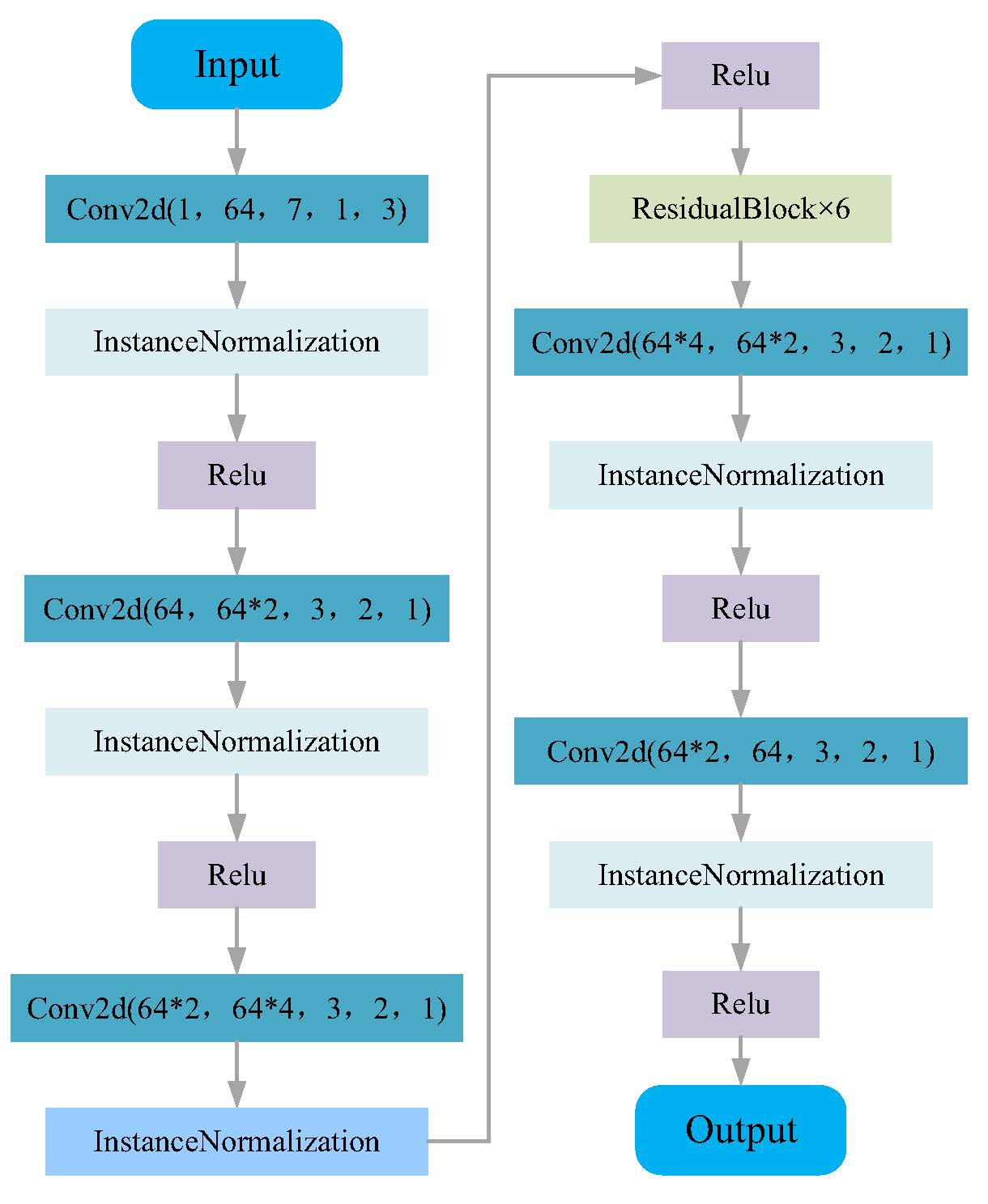}
   \end{tabular}
   \end{center}
    \caption{The figure displays the structure of the neural network proposed in this paper, which incorporates multiple convolutional neural networks and residual blocks to efficiently encode the features of images.
   \label{network structure}}
   \end{figure}

We propose to build the autoencoder with convolutional layers and residual blocks as shown in figure \ref{network structure}, since this structure is better in processing images. The structure includes Convolutional layers (Conv2d), Instance Normalization layers (InstanceNormalization), Relu layers (Relu), Transpose Convolutional layers (ConvTranspose2d) and Residual blocks (Residual Block). The number of Residual blocks \citep{zagoruyko2016wide} could be arbitrary to increase representation abilities of the  neural network. After convolutional layers, we propose to use instance normalization to improve the generalization ability of neural networks. This structure ensures faster convergence while ensuring the preservation of image features and the integration of global information during training, which is beneficial to image feature learning.\\

According to real application requirements, we choose images of high quality in the same observation task as reference images. Then we train the autoencoder using these images and employ the mean square error (MSE) as the loss function. This loss function compares the reconstructed images with the input images, as illustrated below:
\begin{equation} \label{eq1}
MSE = \frac{1}{M \times N}\sum_{0 \le i <  N} \sum_{0 \le j <  M} (f_{ij} - f'_{ij})^2 ,
\end{equation}
where $M$ and $N$ stand for size of images in horizental and vertical directions, $f_{ij}$ and $f'_{ij}$ stand for grey scale values for pixel $i,j$. In fact, the mean square error tends to amplify the impact of outliers as a result of the squaring operation. By training the autoencoder with the mean square error as the loss function, we are able to capture the intricate details of high-resolution images within the autoencoder. Consequently, the autoencoder is capable of reconstructing any given image within the feature space defined by high-resolution image features. The disparity between the reconstructed images and the original images serves as an indicator of the deviation between the input images and the high-resolution images, enabling us to assess the quality of the images. Further elaboration on the specifics of our algorithm will be provided in the subsequent subsection.\\

\subsection{The Workflow of the Framework} \label{sec:rotate}
With the autoencoder for image quality evaluation, we are able to build the framework for image quality evaluation and masking. The framework consists of the following distinct parts:\\
\begin{itemize}
\item Reference Image Selection Step: The selection of reference images plays a crucial role in determining the tolerance of our algorithm. If we choose images with excellent seeing conditions and low background noise levels, even moderately blurred images may be considered as low-quality. On the contrary, selecting images of medium to high quality allows the algorithm to handle different levels of blur and noise more effectively. Additionally, the reference images should include celestial objects typically found in actual observation images, such as stars and galaxies. Typically, we select several hundred reference images of acceptable quality that include different types of celestial objects.
\item  Image Preprocessing Step: Both the reference images and the images to be evaluated undergo the preprocessing step. In this step, the images are divided into small patches and each patch is normalized to a matrix with floating-point values between 0 and 1. The patch size is an important parameter for the algorithm. If the patches are too small, the features extracted may not effectively evaluate the images, while excessively large patches would result in oversized masks for subsequent processing. In this paper, we present results obtained from both large and small patches, tailored to the specific requirements of the application for readers' reference.
\item  Data Augmentation and training Step: Since the number of reference images is limited, we employ data augmentation techniques to generate additional reference images to increase the complexity and diversity of the data. By randomly cropping, flipping, and rotating the reference images, we create augmented versions that can enhance the training process. The neural network is trained using the selected reference images. The MSE defined in equation~\ref{eq2} is utilized as a reference metric during the training process.
\item  Deployment Step: Once the training is complete, the images to be evaluated are fed into the neural network, which reconstructs new images based on the learned features. To better reflect variations, the mean absolute error (MAE) between the reconstructed images and the original images is then calculated to evaluate the image quality, 
\begin{equation} \label{eq2}
MAE = \frac{1}{M \times N}\sum_{0 \le i <  N} \sum_{0 \le j <  M} \|f_{ij} - f'_{ij}\|.
\end{equation}
Patches of images with lower MAE are considered to have higher quality. When the MAE of a patch falls below the threshold, the framework will mask that part of the image. For different data sets and task requirements, it is necessary to manually set different thresholds.
\end{itemize}

\section{Performance Evaluation of the Framework} \label{sec:3}
In this section, we will assess the efficacy of our approach through tests conducted on both simulated and real observational images in two different application scenarios. To generate simulated observational data obtained from ground-based telescopes, we utilize the Sky Maker and the Stuff \citep{bertin2009skymaker, bertin2010stuff}. These tools simulate the impact of diverse factors, including varying levels of complex background noise and point spread functions (PSF) with different full-width half-magnitudes (FWHM). Armed with the ground truth values of FWHM and complex background noise levels, we can quantitatively describe the robustness and effectiveness of our method under different conditions. Additionally, we integrate observational data collected by the Ground Wide Angle Camera Array (GWAC) \citep{wei2016deep}. This telescope array comprises various telescopes designed to automatically observe time-domain astronomical events. Data obtained from the GWAC may be influenced by sky backgrounds, exposure failures, and other effects. To tackle this challenge, we have implemented the framework proposed in this paper to process these images, successfully concealing such artifacts. We employ our algorithm to assess simulated images based on their PSFs and background levels. Meanwhile, our algorithm is utilized to assess real observation images, identifying regions substantially impacted by background noise. Further discussions on these processes are provided in subsequent sections.\\
 
\subsection{Performance Evaluation Using Simulated Images with PSFs of Varying FWHMs} \label{sec:3.1}
\begin{figure} [ht]
   \begin{center}
   \begin{tabular}{c} 
   \includegraphics[height=4cm]{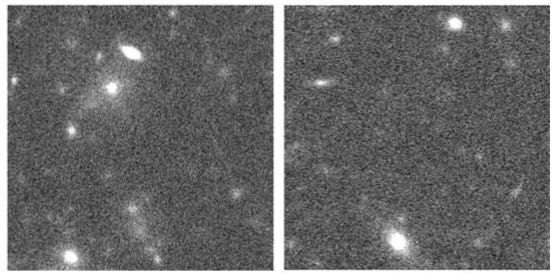}
   \end{tabular}
   \end{center}
   \caption{The figures presented here depict simulated images generated based on the previously defined parameters. As illustrated, these images encompass a variety of celestial objects, including stars and galaxies.
   \label{simulated images}}
   \end{figure}

In this part, we firstly generate a catalogue which contains stars and galaxies with default parameters defined in the Stuff \citep{bertin2010stuff}. Then we simulate images obtained by a telescope with the SkyMaker \citep{bertin2009skymaker}. The telescope has a diameter of 2.5 meter and a pixel scale of observation images is 0.2267 arcsec. The observation is carried out in the V band and the sky background noise level is 20.0 mag/$arcsec^2$. The exposure time is 60 seconds. The read out noise level is 2 $e^{-}$ and the dark current level is 1 $e^{-}/s$. With parameters defined above, we modify the FWHM of PSFs to generate images with different blur levels. Two frames of such images are shown in figure \ref{simulated images}. As shown in this figure, the simulation images contain galaxies, stars and realistic noises, which could well represent observation images.\\

\begin{figure} [ht]
   \begin{center}
   \begin{tabular}{c} 
   \includegraphics[height=5.5cm]{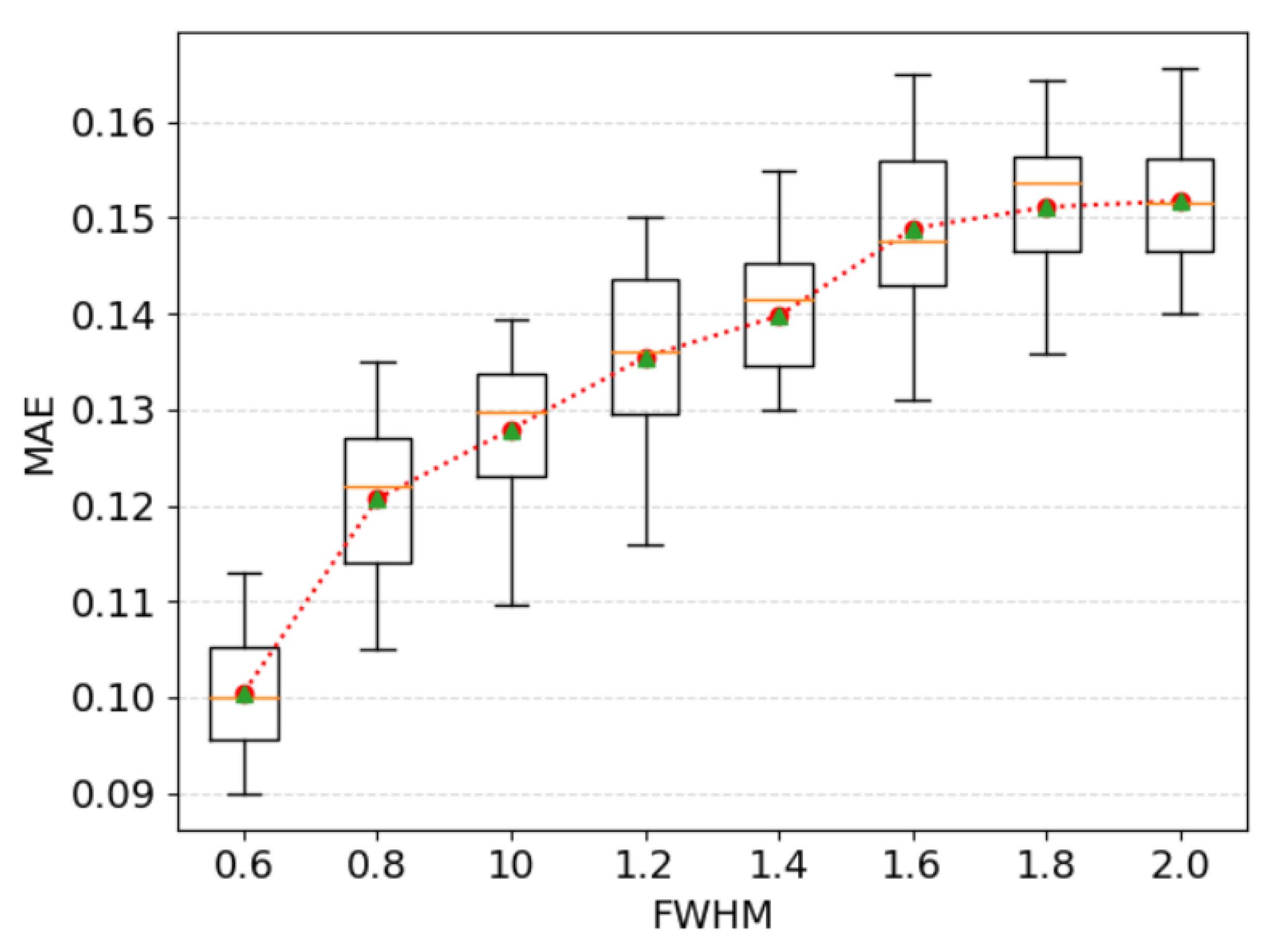}
   \end{tabular}
   \end{center}
   \caption{This figure shows MAE of new sets of simulated images obtained by our methods (from FWHM = 0.5 to FWHM = 2.0). The black straight line stands for mean values of all images and the red dotted line stands for MAEs of the same image with different blur levels.}
   \label{Box plot of simulated image}
   \end{figure}

We have carefully selected a range of reference images with varying FWHM values (from 0.5 $arcsec$ to 0.8 $arcsec$) to train the neural network. These reference images serve as a benchmark for images with desirable qualities. Once the training is completed, the encoder learns the distinguishing features of high-quality images, enabling it to assess the quality of other images. To evaluate the performance of our method, we first assess the quality of images with different FWHM values. We generate additional images with FWHM ranging from 0.5 arcsec to 2.0 arcsec. These images incorporate several new star catalogues, which provides a more rigorous test of our method's robustness. With these images, we compute the mean absolute error (MAE) between these images and their reconstructed counterparts using the approach discussed in Section~\ref{sec:rotate}. The results are illustrated in Figure~\ref{Box plot of simulated image} as box plot in black colour. As evident in the figure, our method demonstrates robustness in evaluating images with varying FWHM values, although the presence of different contents introduces some variability in the box plot. Furthermore, Figure~\ref{Box plot of simulated image} illustrates the MAE values for the same image with PSF of different FWHM values (in red dotted line), demonstrating a monotonous change that indicates the effectiveness of our method.\\

\subsection{Performance Evaluation Using Simulated Images with Complex Background Noises} \label{sec:3.2}
\begin{figure} [ht]
   \begin{center}
   \begin{tabular}{c} 
   \includegraphics[height=13cm]{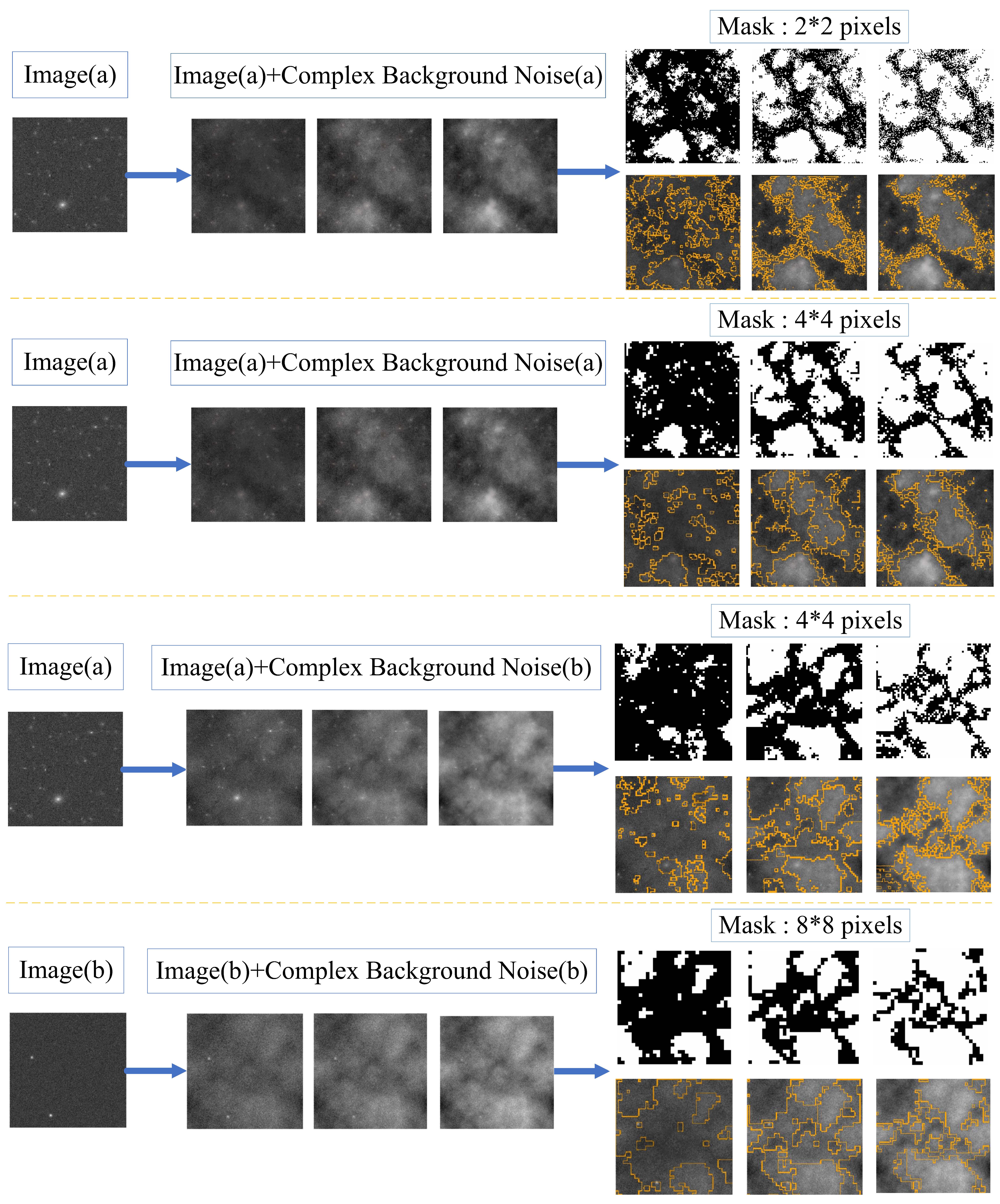}
   \end{tabular}
   \end{center}
   \caption{The figure shows simulated images with different levels of complex background noise. The second and third parts show the results of the same original simulated image using different levels of complex background noise. The first column image represents the original simulated image, while the second to fourth columns images represent images with different complex background noise. Images of columns 5th to 7th are results obtained by our method with different mask sizes and combinations of original image and the contour of ground-truth complex background noise distribution.}
   \label{noise image with mask}
   \end{figure}

In this section, we evaluate the performance of our framework using simulated images that incorporate varying levels of background noise originating from clouds, the sky background, and diffraction spikes caused by bright stars. The presence of such background noise introduces potential risks to the photometric accuracy and may lead to false detection results. To quantitatively assess the effectiveness of our method, we use Skymaker and Stuff software to generate simulated images, following the default parameters outlined in Section~\ref{sec:3.1}. Additionally, we simulate the impact of complex background noise on these images, introducing large-scale nonuniformities. We adopt the approach proposed by \citet{jia2015simulation, jia2022digital} to generate the distribution of complex background noise. The gray scale values of these phase screens are normalized to a range of 0 to 1, representing the coverage levels of complex background noise. In the simulated images, complex background noise obstructs light from celestial objects and generates a light distribution, leading to significant areas of contamination. The reflection light are calculated according to gray scale values of the phase screens and the peak values are set as a random variables. The images in Figure~\ref{noise image with mask}, from the second to the fourth column, represent images with varying levels of complex background noise.\\
   
\begin{figure} [ht]
   \begin{center}
   \begin{tabular}{c} 
   \includegraphics[height=6cm]{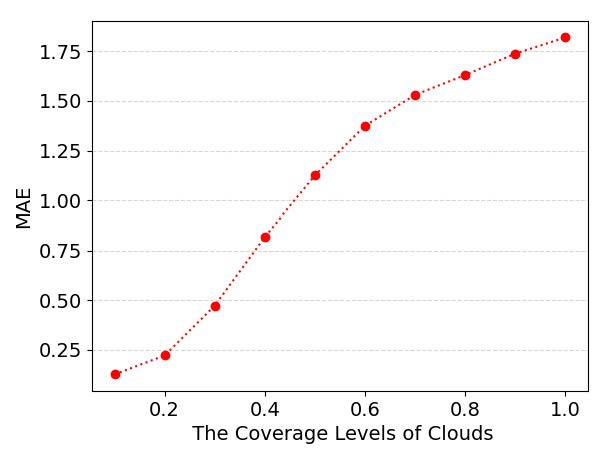}
   \end{tabular}
   \end{center}
   \caption{This figure shows MAE and complex background noise. As this diagram shows, the MAE could be used to assess complex background noise levels. However, when the complex background noise level is high, our algorithm is not sensitive to changes. }
   \label{MAE of different background noise}
   \end{figure}

We begin by assessing the efficacy of our framework in evaluating image quality. We multiply the complex background noise coverage from 0.1 to 1.0 to obtain images with different complex background noise. Then, we use the model to evaluate the quality scores of simulated images influenced by different complex background noise levels. Figure~\ref{MAE of different background noise} presents the MAE between the reconstructed and original images as a function of complex background noise. The figure illustrates that our method reliably captures the image quality as complex background noise increases. However, it should be noted that the MAE becomes less sensitive to complex background noise when complex background noise reaches high levels.\\

In real observations, it is necessary to mask images containing complex background noise and process the remaining portion to derive scientific outcomes. By leveraging the MAE as an indicator of image quality, we can segment the original images into smaller patches and assess the MAE for each patch. By setting a user-defined threshold, we differentiate between images with and without complex background noise, resulting in masked images that specifically highlight regions with complex background noise. The threshold value should be carefully chosen to meet the user's requirements. A higher threshold may inadvertently misclassify some celestial object images as complex background noise, whereas a lower threshold may erroneously identify complex background noise areas as celestial objects. It should be noted here that the threshold is directly related to the relative strength between the light brought by the complex background noise. In this paper, we recommend to set the same value as the threshold for all images whose sky background does not change significantly. In this study, we have set the threshold at 1.25 for all images. The size of patches to be masked is another parameter that users can customize. Larger patch sizes facilitate more accurate feature extraction and recognition of complex background noise but come at the cost of increased mask size. On the contrary, smaller patch sizes produce finer masks but may compromise accuracy in complex background noise recognition. In Figure~\ref{noise image with mask}, we demonstrate the masking results using patch sizes of $2\times 2$, $4\times 4$, and $8\times 8$ pixels. As illustrated in the figure, different patch sizes produce varying masking results. In practical applications, employing patch sizes that are at least two times larger than the size of FWHM of PSFs, would yield improved results. In the next section, we will show the real applications of our algorithm with the above parameters.\\

\subsection{Performance Evaluation Using Real Observation Images from the GWAC}
Building upon the results obtained in earlier sections, we now apply our method to real observational images obtained by the GWAC. This unique instrument employs multiple telescopes and cameras to collect white-light time-domain astronomical observation images, enabling the monitoring and analysis of time-domain astronomical events. The wide field of view and rapid response capabilities of the GWAC make it ideal for studying explosive astronomical events such as Gamma-Ray Bursts (GRBs), supernovae, and flares from stars \citep{Xu2020THEGD, li2023whitelight, xin2023prompt}. The GWAC monitors the sky with a cadence of 15 seconds, comprising 10 seconds for exposure and 5 seconds for readout. This setup is designed to detect magnitude variations of celestial objects across the entire night swiftly, facilitating prompt identification and monitoring of astronomical transients. Covering a vast area of 5000 square degrees in the sky, the complete camera array operates throughout the night, with each camera capturing up to 2400 images per night at a resolution of $4k \times 4k$. A major challenge in processing images from the GWAC is the impact of complex backgrounds, arising from lunar presence, mechanical shutter malfunctions \citep{Kri_i_nas_2023} or other effects. The complex background introduces gradient to magnitude of celestial objects, leading to increased photometric errors and lower detection efficiency for faint stars. Consequently, accurately assessing the extent and severity of image contamination by complex background becomes crucial. Therefore, we propose to use our framework to process these images.\\

The images captured by the GWAC, initially sized at $4196\times 4136$ pixels, undergo cropping to dimensions of $4096\times 4096$ pixels to eliminate non-informative black borders. Following this, they are subdivided into smaller patches measuring $512\times 512$ pixels in this study to alleviate the demand on GPU memory. In real applications,the original observation images can undergo processing using GPUs with larger memory capacities. For instance, our framework requires approximately 1320 seconds to process 600 images sized at $4096\times 4096$ pixels on an RTX 3090 GPU, averaging 2.2 seconds per image when divided into patches measuring $512\times 512$ pixels. Utilizing a computer equipped with a GPU boasting larger memory, such as the Nvidia A100, would result in a processing time of less than 0.034 seconds. To train our framework effectively, we select only pristine images for the training set, prioritizing those captured on clear moonless nights without shutter shadowing. Once trained, the framework processes GWAC images by performing patch-based evaluation. We assess the reconstructed patches against their originals with the MAE. Averaging these patch scores yields a quality score for the entire image, effectively flagging those heavily affected by complex background noise. Subsequently, these flagged images are diverted to a dedicated thread within the pipeline for specialized processing using alternate approaches. Two different tests are conducted in the following parts to show effectiveness of our method.\\

\begin{figure} [ht]
   \begin{center}
   \begin{tabular}{c} 
   \includegraphics[height=5cm]{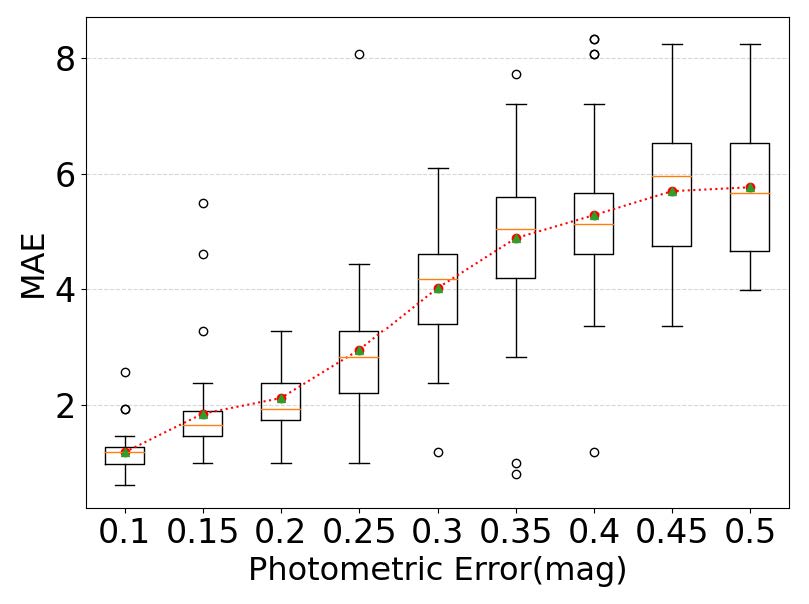}
   \end{tabular}
   \end{center}
   \caption{The figure shows the relationship between the MAE and the photometric errors. The black straight line stands for mean values and the red dotted line stands for the median values of all images.}
   \label{Box plot of GWAC}
   \end{figure}

In the first test, we have randomly selected 600 images from the GWAC dataset covering the period from 2018 to 2021 and have divided them into 38,400 image patches sized at $512\times 512$ pixels each. Subsequently, our framework has been utilized to compute the MAE for each of these patches. Then, we have performed aperture photometry in these images. The photometry results are then matched with the Gaia DR3 catalog for magnitude calibration. Finally, we have calculated the difference between our photometry results and the G magnitudes provided by Gaia. Visualization of the results is presented in Figure~\ref{Box plot of GWAC}. As evident from the figure, as the MAE increases the photometric error also increases, demonstrating the ability of our model to effectively capture the influence of background variations.\\

\begin{figure} [ht]
   \begin{center}
   \begin{tabular}{c} 
   \includegraphics[height=7cm]{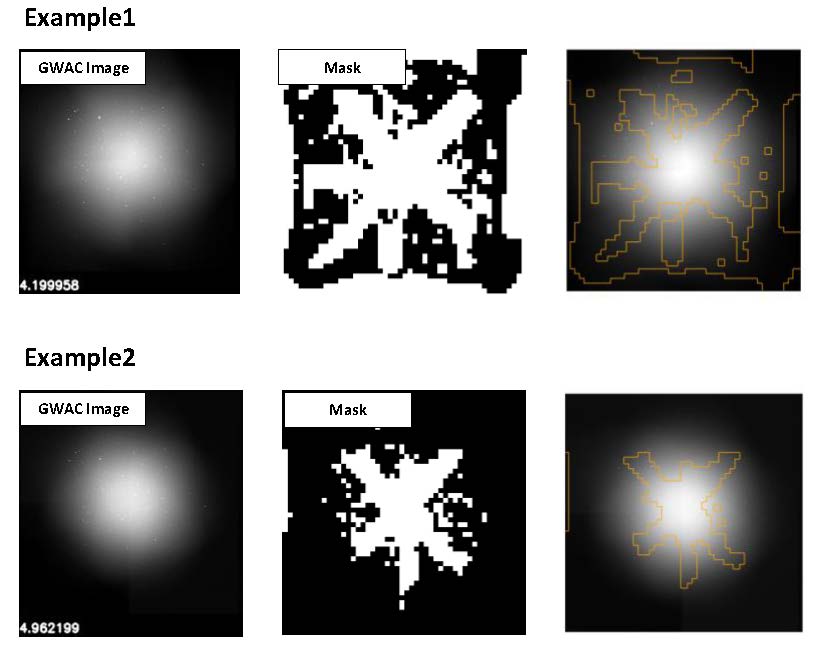}
   \end{tabular}
   \end{center}
   \caption{Two frames of GWAC images affected by shutter malfunctions and their corresponding masks. }
   \label{GWAC image with mask}
   \end{figure}

Leveraging the calculated MAE values, we can generate masks for patches significantly impacted by background noises. These masks allow us to apply targeted processing to these areas, mitigating photometry errors and false detection rates. This ultimately leads to improved accuracy in light curve classification and transient detection. Therefore, we adopt a similar approach as outlined in Section 3.2, setting a threshold of 3.5 for mask generation. However, it is crucial to acknowledge that bright stars can introduce isolated patches with relatively high MAE values. To address this, we perform dilation and erosion operations when overlaying the mask on the original image. As illustrated in Figure~\ref{GWAC image with mask}, segmenting the original image into smaller patches and utilizing MAE as a quality metric effectively identifies patches affected by background noises introduced by shutter malfunctions.\\                                                                                            
   
In the second test, we have selected a sequence of images from the GWAC dataset. Given the high temporal resolution of the GWAC, we can extract photometry results from these images to generate light curves for detecting transient phenomena, such as variable stars and eclipsing binaries. Obtaining a reliable light curve involves several steps, including source extraction and flux calibration. However, complex background noise in the images may lead to false signals. Typically, techniques such as binarization and median filtering are employed to process light curves and mitigate photometry errors \citep{chen2023meteor}. The binarization threshold is usually set as twice of the average pixel standard deviation of the image. Nonetheless, there is a risk of missing important astronomical transients or introducing excessive noise due to variations in backgrounds. With our method, contaminated background areas can be swiftly identified and filtered out, allowing preprocessing operations to focus on localized regions. Subsequent sections will present the performance of our framework in detail.\\  

\begin{figure} [ht]
   \begin{center}
   \begin{tabular}{c} 
   \includegraphics[height=7cm]{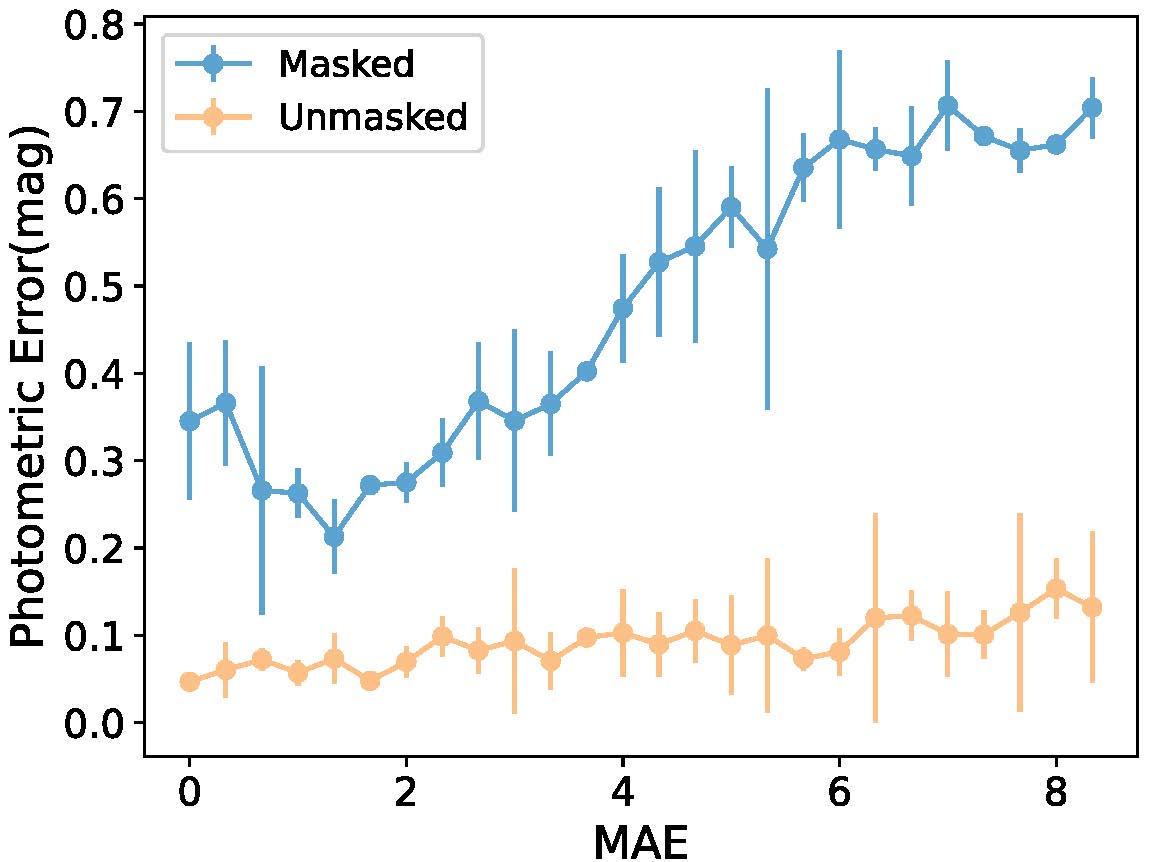}
   \end{tabular}
   \end{center}
   \caption{The figure shows the photometric error comparison between the masked region and other areas for each image under different MAE results.}
   \label{Performance of mask algorithm on GWAC-1}
   \end{figure}

We have conducted an assessment using 600 continuous observation images, applying our method to evaluate the MAE of these images and derive masks accordingly. Within these images, we have identified and selected 80,000 celestial objects, calculating the photometric error of them. Based on their respective positions, we have categorized these stars as either masked or unmasked. "Masked" stars denote those whose photometry results are impacted by background noise, while "unmasked" stars indicate those unaffected by such noise. The relationship between photometric error and MAE values is depicted in Figure~\ref{Performance of mask algorithm on GWAC-1}. When background noise influence is small, there are less discernible differences in photometric errors. However, as background noise influence gradually increases, noticeable disparities in photometric errors between masked and unmasked regions become apparent. These findings demonstrate the efficacy of our masking algorithm in distinguishing areas of image contamination, thereby aiding in the selection of high-quality data during the preprocessing stage.\\  
                                                         
\begin{figure} [ht]
   \begin{center}
   \begin{tabular}{c} 
   \includegraphics[height=7cm]{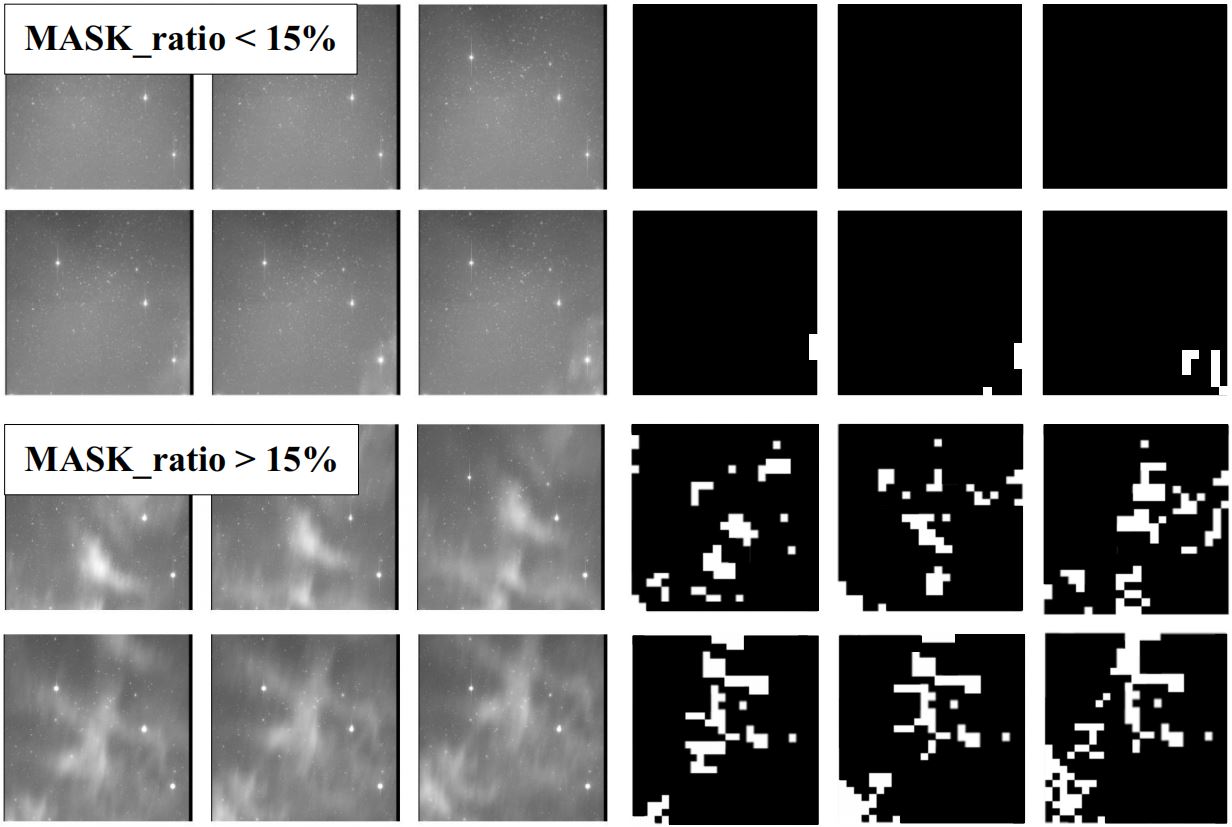}
   \end{tabular}
   \end{center} 
   \caption{The figure shows the data selected by our algorithm with different mask ratios.}
   \label{GWAC_image_with_mask-2}
   \end{figure}

Upon further examination of the images, we have found that passing clouds have affected the observations. Specifically, out of all the observation images, 120 images have been impacted by passing clouds during the 30-minute observation period. We define the mask ratio as percentage of masked regions to the whole image. Among these affected images, 35 exhibit a mask ratio of less than 15\%, while the remaining images have a mask ratio exceeding 15\%, as illustrated in Figure~\ref{GWAC_image_with_mask-2}. Our method could successfully identify these regions. Among the images with a mask ratio exceeding 15\%, photometric errors range from 0.12 to 0.25. Conversely, images with larger mask areas exhibit photometric errors ranging from 0.50 to 0.73.\\

\begin{figure} [ht]
   \begin{center}
   \begin{tabular}{c} 
   \includegraphics[height=7cm]{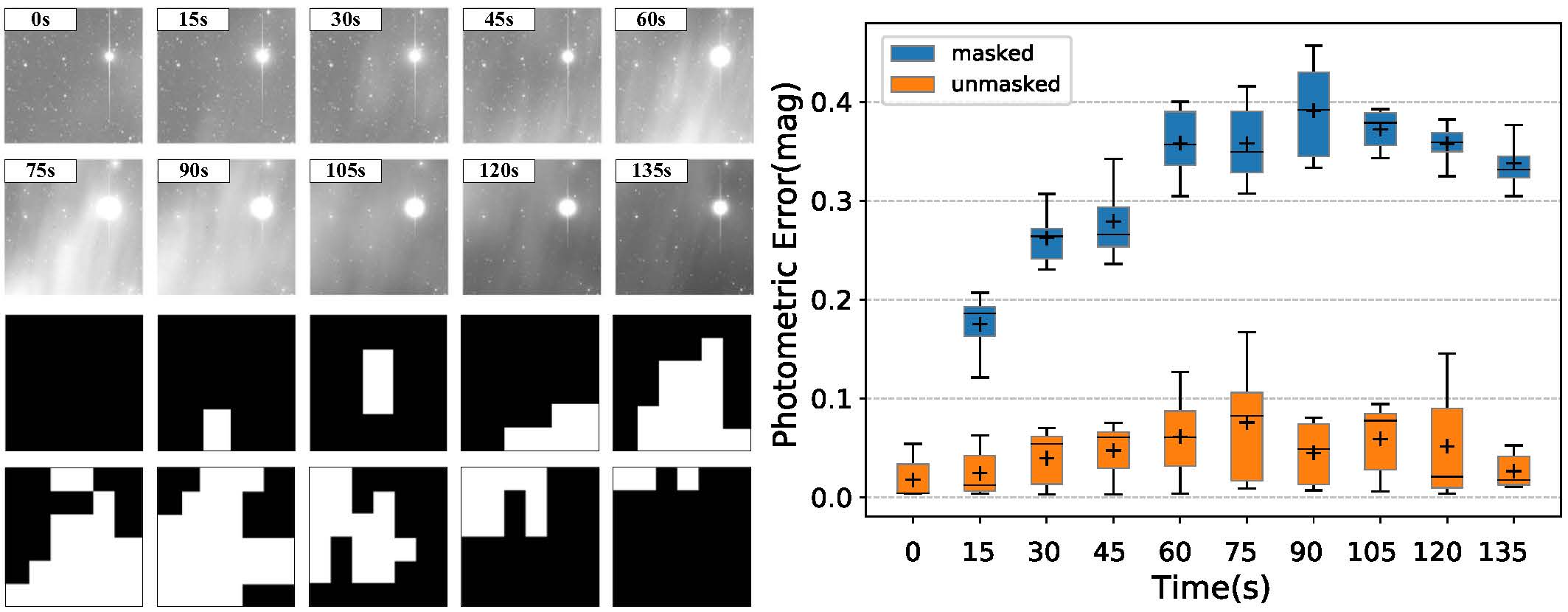}
   \end{tabular}
   \end{center} 
   \caption{The figure shows the masking results by our method in continuous observations, along with the corresponding variations in photometric errors in different regions.}
   \label{Performance of mask algorithm on GWAC-2}
   \end{figure}

We have identified a specific region within the images that is initially clear but later affected by passing clouds. 228 celestial objects with sufficient SNR (magnitude ranging from 9 to 14) are selected as reference stars, while variable stars are excluded from them. These stars are further categorized into masked and unmasked categories according to the masks generated by our algorithm, and photometry is conducted accordingly. The images, along with their corresponding masks and photometry results, are presented in Figure~\ref{Performance of mask algorithm on GWAC-2}. Our algorithm has effectively identified areas with significant photometric errors, resulting in an average improvement of 0.51 magnitudes in photometric error correction. Utilizing local masking algorithms enabled us to maintain photometry errors at low levels, thereby aiding in the reduction of false alarms in subsequent scientific analyses.\\

\section{conclusion}
In this paper, we propose an effective method to evaluate the qualities of images. With several images of high quality as references, we could encode features of these images automatically using an autoencoder. Then images to be evaluated are reconstructed with the autoencoder, and we could compare the reconstructed images and images to be evaluated to obtain scores to evaluate qualities of images. Through splitting original images into small patches and evaluating MAEs of these images, we could mask images with low quality and process rest images for further processing. Simulated images with different seeing conditions, simulated images with different complex background noise levels and real observation images are used to test the performance of our method. Results show that our method is effective. Our method could be used to quickly evaluate or mask images with low quality for different sky survey projects.\\

There are several aspects that need further investigation. Firstly, the neural network utilized in this paper consists of relatively fewer parameters (around 15 million parameters) compared to state-of-the-art large vision models (around several billion parameters). Exploring the use of larger models specifically designed for astronomical images could potentially enhance the performance of our algorithm. Second, our algorithm requires two sets of parameters to be determined by scientists: the patch size for masking and the masking threshold. Currently, these parameters need to be set through a trial and error approach, which adds complexity to the utilization of our method. We will explore alternative methods to automatically determine these parameters. Third, we will integrate the image quality evaluation method with the source detection algorithm to build an end-to-end pipeline to extract scientific information from observation images.\\

\section*{acknowledgement}
Authors would like to thank the reviewer for his/her kindly suggestions, which greatly improve the quality of this paper. The code used in this paper will be shared in the PaperData Repository powered by the China-VO. This work is supported by the National Natural Science Foundation of China (NSFC) with funding numbers 12173027 and National Key R \& D Program of China (No. 2023YFF0725300). This work is supported by the Young Data Scientist Project of the National Astronomical Data Center. We acknowledge the science research grants from the China Manned Space Project with NO. CMS-CSST-2021-A01 and research grants from the Square Kilometer Array (SKA) Project with NO. 2020SKA0110102.\\


\bibliography{reference}{}
\bibliographystyle{aasjournal}



\end{document}